\documentclass[english,11pt]{article}

\newcommand{\be}{\begin{equation}}
\newcommand{\ee}{\end{equation}}
\newcommand{\rf}[1]{(\ref{eq:#1})}

\newcommand{\G}{{\cal G}}

\begin{document}

\title{On the Topological Nature of  the Cosmological Constant}
\author{M. D. MAIA\\
Instute of  Physics,
Universidade  de  Brasilia\\ 70910-97- Brasilia D.F.  Brazil\\maia@unb.br
}

\maketitle

\abstract{It is  shown   that  topological  changes  in  space-time are  necessary to  make  General  Relativity compatible  with the Newtonian  limit  and  to   solve the  hierarchy of  the  fundamental interactions. We detail how topology  and  topological  changes appear in General Relativity and  how  it  leaves  an  observable  footprint in  space-time. In  cosmology  we  show  that such topological  observable  is the  cosmic  radiation  produced by the   acceleration of  the universe. The  cosmological constant  is      a very particular case which occurs  when the  expansion of the universe into the vacuum occurs  only  in the   direction of the cosmic time flow.

}

\section{The Cosmological Constant  Problem}

The  cosmological  constant $\Lambda$  was
introduced   by  Einstein in  1917 to  implement his  static universe, but  he  removed  it soon  after,  in face of the  evidences for an expanding universe and
 the understanding that the presence of  $\Lambda$ prevents the  emergence of   the Minkowski   space-time as  a particular  solution of  Einstein's  gravitational  equations.
From the mathematical point of  view,  the cosmological  constant  originated   from  the contracted Bianchi  identity  $(R_{\mu\nu} -\frac{1}{2}R g_{\mu\nu})_{;\rho} \equiv 0$.  This   conveys  to  the   conclusion that  the Einstein tensor  (between the parenthesis)   should  be  proportional to   $(g_{\mu\nu})_{;\rho} \equiv 0$  which is  the    defining  condition  for  the  Levi-Civita  connection.
Thus,   $g_{\mu\nu}$  behaves  as   a  constant  under the metric covariant derivative.   Since  this same condition  was  already  used  for  the  derivation of  the Einstein  tensor,   the  identification  of  the  contracted  Bianchi  identity  to  something proportional  to $g_{\mu\nu}$ is  a mathematical redundancy,  using an  argument that has  been  already spent.

Yet,  the  cosmological  constant  $\Lambda$  has been   claimed  to be the  simplest  explanation for the    acceleration   of  the  universe within   the  $\Lambda$CDM  paradigm.
This  follows  from the  derivation  of the   vacuum  energy  density  of  quantum  fields \cite{Zeldowich} as a  solution of  the  semiclassical vacuum Einstein's  equations\footnote{We  use the  following  notation: Greek  indices  run  from 1  to  4,  small case  Latin indices  run from  5 to $D$  and  large  Latin indices  run   from  1  to  $D$  and  .}
\[
R_{\mu\nu}- \frac{1}{2}R g_{\mu\nu}  -\Lambda g_{\mu\nu} =-8\pi G  <\rho_v> g_{\mu\nu}
\]
where     $<\rho_v>$   denotes  the constant  vacuum  energy  density. Since  the  evaluation of  the  energy  density is  made  with the standard  quantum  field  theory  defined in Minkowski's  space-time, then it must cancel exactly  with $\Lambda$:
\be
<\rho_v> =  \Lambda/8\pi G
\ee
so that   the existence of the  Minkowski space-time as  a  solution of  the classical  Einstein's  equations would  be  guaranteed.
However, the theoretical estimates of  $<\rho_v>$  gives $<\rho_v>  \approx 10^{76}  GeV^2/c^4$.
On the  other hand,  the acceleration of  the universe   indicates  an  experimental value  of $ \Lambda/8\pi G \approx  10^{-47} GeV^2/c^4 $. This   very  large  difference  cannot be   fixed within quantum field theory  \cite{Weinberg,Weinberg1}.
Consequently    either  from the geometrical  or from  the physical  points of  view, the  presently accepted  explanation  for the acceleration of the universe is  ill  founded.

\section{Topology  of Space-times}

Perhaps the least  acknowledged  fact about  Einstein's  gravitational theory is  that it  requires  the  notion   of  topological  change to make it   consistent with the presence of Newton's  gravitational  constant  $G$ in  Einstein's  equations
\be
R_{\mu\nu}-\frac{1}{2}R g_{\mu\nu} =8\pi G T_{\mu\nu}  \label{eq:EE0}
\ee
Newton's  space-time  is  a  manifold  with  topology   $I\!\!\!\!R^3 \times I\!\!\!\!R$,  meaning  that   the  distances in   the space-like   sections parameterized by $I\!\!\!\!R^3$  are  independent  of  the  open intervals  of  the absolute time  in $I\!\!\!\!R$  as  dictated by   the  Galilean symmetry  of  Newton's  space-time \cite{MaiaGFI}. The  factor  $8\pi$   comes  from the   spherical solid  angle  $4\pi$
defined in the  3-dimensional  simultaneous  sections  appearing in  Poisson's   equation $\nabla^2 \phi =4\pi G \rho$.  Here $\rho $ is  the  3-dimensional mass  density.

On the other  hand,  in Einstein's  General  Relativity  the space-time  manifold is  parameterized by
$I\!\!\!\!R^4$  without a separation between  time  and  space  as  dictated  by the  diffeomorphism invariance of the theory.  Therefore  it has  a   different  topology,  which we call Lorentzian topology.
In this case,  once  a   coordinate is  chosen  to be
the  time, the diffeomorphism  invariance  breaks  down  and  the Lorentz  topology of the  space-time  changes  to the product  topology  $I\!\!\!\!R^3 \times I\!\!\!\!R$,    the same  topology of the Newtonian  space-time,   but with the difference that  the  space  and  time  factors  have  different  meanings \cite{Gibbons}.  Therefore, the  change in topology in  General relativity is  a   necessity  to make   the  appearance  of  $G$  consistent with the  relativistic  principles.
It is also relevant to notice   that  the   value  of   $G$  introduces  an  additional  problem  for General Relativity,  namely  the  hierarchy between gravitation  and  the  other fundamental (gauge) interactions. In terms  of  energy it is  like  the difference  between   $10^{15} $TeV  to  $1$TeV in  scales of  energy,  with nothing  happening in between.

The  gauge  interactions start  with
 Maxwell's equations   written in the  form
\be
 D_\nu F^{\mu\nu}  =4\pi j^\mu ,\;\;\;
 D_\nu  F^{*\mu\nu} = 0
\ee
where  $F*$  denotes  the Hodge dual  of  $F$;  $D_\mu  =\partial_\mu  +  A_\mu$  and $F_{\mu\nu} = [D_\mu , D_\nu ]$.  $A$ is the gauge vector-potential to be  obtained  by  solving the  above  equations.
Such  structure  is also  common to  the  weak and  strong  nuclear  forces. The  first of  these  equations tells  that the  3-form or  covariant antisymmetric tensor   in the left  hand  side must  equal  to the  current one  form or  covariant vector  in the  right  hand  side.  This  is possible  only  in  four  dimensions  and  consequently, the  existence of  the  gauge  fundamental  interactions  are  consistent only in  four-dimensional  space-times.

Einstein's  gravitation is  not  a  gauge theory  for  several reasons:   it  does not  have  the same  dual property;  the  main  variable  of  a gauge theory is  the connection $A$  determined in the  adjoint representation of the gauge  group,   whereas  in  Einstein's  gravity  the  connection is  postulated.  Finally, the   Einstein-Hilbert action is  linear  in the Ricci  scalar, while  the  actions of   gauge  fields  are  quadratic in the  curvature. Thus, in Einstein's  theory,   the gravitational  field  does not possess the  same four-dimensional constraint,  meaning that   gravitation   may  propagate along  extra dimensions in excess of four,  but   it also propagate  in a  space-time   subspace  of  some higher-dimensional  space.

In  spite of  such  conceptual  differences, the   co-existence  between gravitation and  gauge  fields
(and  even  their  unification)  can  be  solved  without  modifying  General  Relativity.
Indeed,  using  the  same previous  arguments to  justify   the presence of  $G$  in  Einstein's  equations,  the coupling constant between  gravitation and  matter   may  vary  depending  on the number   $N$ of  extra  dimensions   accessible  by  gravitation.
Suppose  we have  a   space-time  $V_4$  embedded in a  Riemannian geometry of $D =4 + N$  dimensions  with the Lorentz topology  $I\!\!\!\!R^{D}$.  By  fixing  $N$  extra  dimensions,   that topology  changes  to the product  topology   $I\!\!\!\!R^{4} \times I\!\!\!\!R^N$.  Since the total number of space dimensions  accessible by  gravitation is  now greater  than three,  the    gravitational constant  cannot be  the same  as  the  Newtonian  $G$.  The  new  value  $G^*$ must be established  experimentally, in  cosmology and  in  high energy    experiments \cite{ADD}.

 Admitting that the  definition of  the  metric  is  the  same  for  both spaces (an isometric  embedding), the  metric of the  embedding  space  must  also  follow from the Einstein-Hilbert principle
\be
\frac{1}{G_*}\frac{\delta}{\delta {\cal G}_{AB}}   \int{^D{\cal R}\sqrt {\cal G}d^D v}  =0  \label{eq:EH}
\ee
where     $^D{\cal R}$  denotes  the higher-dimensional  scalar  curvature.  We  Obtain  the  D-dimensional  Einstein's  equations
\be
  R_{AB} -\frac{1}{2}R {\cal G}_{AB}  =G_* T_{AB},\;\;  A,B=1..D  \label{eq:bulkEE}
\ee
where   $T_{AB}$   denotes  the  energy-momentum tensor  of the known material  sources.
The projection of these  equations  in the  four-dimensional  space-times is  obtained  simply  by writing them  in the   Gaussian  coordinate  system  defined  by the embedded  space-time  and the  $D-4$ unit normal vector fields. We  obtain in general three  sets of  equations involving  the  metric  $g_{\mu\nu}$,  the  extrinsic  curvatures $k_{\mu\nu a}$  and  the  third fundamental form (or  torsion vector) $A_{\mu  ab}$:
\begin{eqnarray}
R_{\mu\nu}-\frac{1}{2}R g_{\mu\nu}-
Q_{\mu\nu}
&&=G  T_{\mu\nu} \label{eq:BE1}\\
\phantom{x}\hspace{-14mm}
k_{a\mu ;\rho}^{\rho}\! -\!h_{a,\mu} \!+\! A_{\rho c a}k^{\rho \;c}_{\;\mu}\! -\!A_{\mu c a}h^{c}
&&= -2\alpha_* (T^*_{\mu a} -\frac{1}{N+2}T^* g_{\mu a})\label{eq:BE2}\\
S_{ab} -Sg_{ab}- \frac{1}{2}[R-K^{2} +h^{2}]g_{ab} &&=\alpha_* T^*_{ab}
\label{eq:BE3}
\end{eqnarray}
where  we have  denoted   $S_{ab} = {\cal R}_{AB}\eta^A_a\eta^B_b$ and  $S=g^{ab}S_{ab}$  and
\be
 Q_{\mu\nu}  =  g^{ab}k^{\rho}{}_{\mu
a}k_{\rho\nu b}-g^{ab}h_{a}k_{\mu\nu
b}-\frac{1}{2}(K^{2}-h^{2})g_{\mu\nu}  \label{eq:Q}
\ee
It can be   seen  by  direct  calculation that  $Q^{\mu\nu}{}_{;\nu}=0$,  so that   $Q_{\mu\nu}$ can be associated  with an   observable in  space-time.

The  embedding itself is  given by a map  $X:  V_{4}  \rightarrow  V_D$,  satisfying  the  embedding  equations
$X^A{}_{,\mu} X^B{}_{,\nu}\G_{AB}= g_{\mu\nu}, \;
X^A{}_{,\mu}\eta^B \G_{AB}=0, \;
\eta^A \eta^B \G_{AB}=1$,  where  $\eta_a^A$,  $a=1..N$ are the  components of  the  $N$  unit normal  vectors  to the space-time. The  variations of  these  normals when their  foot  are  displaced  in  $V_4$ are  expressed  in terms  of two    coefficients, the  extrinsic  curvature  $k_{\mu\nu a}$ and  the  torsion vector $A_{\mu ab}$  appearing in
\be
\eta^A_{a,\mu}=  -k^{\nu}_{a\mu}   X^A_{,\nu} +A^b_{\mu a} \eta^A_b, \;\; \; k_{\mu\nu a} =k_{\nu\mu a},\;\;\; A_{\mu ab}=-A_{\mu ba}
\ee
Together,  $g_{\mu\nu}$,  $k_{\nu\mu a}$ and  $A_{\mu ab}$,  define the  geometry and  topology of  the  space-time.  The  integrability  conditions  for  the  above equations   are  given by the  components  of the Riemann tensor  of the embedding  space evaluated  in the  Gaussian frame of  $V_4$,  known as  the Gauss-Codazzi-Ricci  equations \cite{Eisenhart}.
 \begin{eqnarray}
&&\phantom{x}\hspace{-1,5cm}^D{\cal
R}_{ABCD}{\cal Z}^{A}_{,\alpha}
{\cal Z}^{B}_{,\beta}{\cal Z}^{C}_{,\gamma}{\cal Z}^{D}_{,\delta} =R_{\alpha\beta\gamma\delta} -
2g^{mn}k_{\alpha[\gamma m}k_{\delta]\beta n}
 \label{eq:Gauss}\\
&&\phantom{x}\hspace{-1,6cm}^D{\cal R}_{ABCD} {\cal
Z}^{A}_{,\alpha} \eta^{B}_{b}{\cal Z}^{C}_{,\gamma}{\cal
Z}^{D}_{,\delta} =k_{\alpha[\gamma b; \delta]} -
g^{mn}A_{[\gamma mb}k_{\alpha\delta]n }\label{eq:Codazzi}\\
&&\phantom{x}\hspace{-1,8cm} ^D{\cal
R}_{ABCD}\eta^{A}_{a}\eta^{B}_{b} {\cal Z}^{C}_{,\gamma} {\cal
Z}^{D}_{,\delta} = -2g^{mn}A_{[\gamma ma}A_{\delta]n b} -2A_{[\gamma a b ; \delta]}
- g^{mn}k_{[\gamma m a}k_{\delta]nb} \label{eq:Ricci}
\end{eqnarray}
These  equations represent  the Frobenius  expression  for  the  curvature  associated  with  a  given  connection  and  they   are  therefore the  integrability  conditions  for  the embedding  and as  such  they are  not dynamical  equations.

Notice  that  in the above  considerations  all principles of  General  Relativity  remain valid,  so that  Einstein's  theory remains  the  same. The   Einstein's  equations are  here  seen as  a  consequence  of the Einstein-Hilbert principle.

\section{Cosmology}
 In the  following,      we  consider the example of  $D=5$  with only one  time  dimension and  $^D{\cal R}_{ABCD}=0$.  This is   sufficient to embed the  standard  model of  the universe as proven by the analysis  of the Gauss-Codazzi  equations.
We also  consider that  all  material  sources of  gravitation are made  of   ordinary  matter, interacting  with  gauge  fields,  so that they are  also confined  to  the space-time. Then  equations  \rf{BE3} is  an  identity   and   \rf{BE1}-\rf{BE2} reduce  to
\begin{eqnarray}
&&R_{\mu\nu}-\frac{1}{2}R g_{\mu\nu}-
Q_{\mu\nu}  \label{eq:BE51}
=G T^*_{\mu\nu} \\
&&k_{\mu a;\rho}^{\rho}\! -\!h_{a,\mu}=0 \label{eq:BE52}
\end{eqnarray}
The  propagation of  gravitation  along the  extra  dimensions  generates  a continuous family of  new  metric  geometries.
To  see  this,  consider a point in  an originally given embedded  space-time  given  by  the coordinate  $X^A$, and  its  (continuous)  propagation on the  extra  dimension  $\eta$  with parameter (coordinate)  $y$. The
curve   determined by  that point and  with  tangent  vector   $\eta$  gives   a new point of the embedding manifold  described by   coordinates $Z^A$.  For  small  displacements $\delta y$,   the   result is  given  by  the  Lie transport   of  $X^A$ \cite{Pirani}
\be
Z^A\!\!  =  X^A +   \delta y \mbox{\pounds}_{\bar{\eta}}X^A  \label{eq:Z}
\ee
Repeating  the same  for all points  in  a  neighborhood of $p$  we   obtain congruence of  curves (the orbits of  $p$),  which  characterizes
a one-parameter  group of  diffeomorphisms  in the  embedding space.  In each of these curves, the   extrinsic  curvature plays the  role  of  the  Frenet acceleration in the theory of  curves in space, orthogonal to  the tangent vector and pointing towards  its  center of  curvature.
These points  define  a new  embedded  manifold  withe metric $G'_{\mu\nu}$  and  extrinsic  curvature  $k'_{\mu\nu}$  provided  they   satisfy  similar    embedding  equations
\be
Z^A{}_{,\mu} Z^B{}_{,\nu}\G_{AB}=g_{\mu\nu},\; \; Z^A{}_{,\mu}\eta^B \G_{AB}=0,  \;\;  \eta^A \eta^B \G_{AB}=1
\ee
Replacing $Z^A_{,\mu}$ in  \rf{Z}
we obtain  the metric  and  extrinsic  curvature  of  the  ``deformed  manifold"  with geometry  given by
\begin{eqnarray*}
&&g'_{\mu\nu} =   {g}_{\mu\nu}-2y {k}_{\mu\nu} + y^2 {g}^{\rho\sigma}{k}_{\mu\rho}{k}_{\nu\sigma}\\
&&k'_{\mu\nu}  ={k}_{\mu\nu}  -2y {g}^{\rho\sigma} {k}_{\mu\rho}{k}_{\nu\sigma}
\end{eqnarray*}
Deriving the  first  equation  with  respect  to  $y$,  we obtain
\be
k'_{\mu\nu}  = - \frac{1}{2}\frac{\partial   g'_{\mu\nu}}{\partial  y}
\ee
indicating that  the gravitational  field defined by \rf{BE51}-\rf{BE52} propagates in the  extra  dimension,  as  specified by the  extrinsic  curvature.

Three particular  solutions of  \rf{BE52} correspond  to  the  cases  where  gravitation  propagates only  in the  four-dimensional space-times:  (1)  When   $k_{\mu\nu}=0$  in   all  directions, then
\rf{BE51}  becomes  identical  to Einstein's  equations  without  topological  considerations. In this case, all     usual   solutions  of  the  intrinsic  General  Relativity  are  reproduced, including the Minkowski  space-time taken  as the  ground  state  of the  gravitational  field. (2)  When   $k_{\mu\nu}=\varphi(x) g_{\mu\nu}$  in  all  directions. In this case,  we obtain  Einstein's  theory  with an  scalar  field $\varphi$. The  space-time  has  an  umbilicus  point  at  $x$.  (3) Finally, if   $k_{\mu\nu}=\alpha_0 g_{\mu\nu}$ where  $\alpha_0$ is  constant,    then   the  space-time is a  space  with constant  curvature \cite{Eisenhart}.  Replacing this  in  \rf{Q},  we obtain  Einstein's  equations  with a  cosmological  constant,  $-\Lambda  =3\alpha^2$,  so that $\Lambda$  appears  as a  very particular   deformation of  the  space-time.

For the  standard  FLRW  cosmology, Friedmann's  equation is  replaced  by \cite{GDEI}
\be
\frac{\dot{a}}{a}^2 + \frac{\kappa}{a^2}=  -\frac{8 \pi G}{3} \rho + \frac{b^2}{a^4}, \;\;\;\mbox{with}\;\;\; \frac{\dot{a}}{a} \propto \frac{\dot{b}}{b} \label{eq:Fried}
\ee
where  we have  denoted $b(t)  = k_{rr}$, the  radial  extrinsic  curvature and  $\kappa$ is the intrinsic curvature  of the space  sections as  given by the sum of  the internal angles  of  a triangle.

 We have  compared this
result  with  the  Turner-White   phenomenological  fluid (x-fluid),  and  later,  using  the  model independent  distance-measurement  statistical  analysis,  we have  found  that  the  term  $b^2/a^4$  fits  the  intervals  compatible  with the  acceleration of  the universe. The  acceleration of the universe  is given  by the radial component of the extrinsic  curvature $k_{rr}$,  and the last  term in  \rf{Fried}  describes the  radiation energy resulting  from the  stretch of  the  geometry as  it  expands  in  the vacuum.

 It is  interesting to notice  that the  space  sections  are  flat in the  sense that $\kappa=0$,  but they  are  not like  flat planes because they  have    non-zero radial  components $k_{rr}$.  The  case  of  $k_{\mu\nu} =\propto g_{\mu\nu}$  corresponds  to the  cosmological  constant.  It  does not occur  here  because  the acceleration would  be exclusively  in the   time  direction and  not  in the  space-sections.
Such  qualitative  difference  may in principle be  detected by  a detailed  analysis  of  the  CMB   power  spectrum.

\end{document}